\documentstyle[12pt]{article}
\begin{document}
\begin{titlepage}
\begin{flushright}
CERN-TH/96-233  \\
hep-th/9609003\\
\end{flushright}

\begin{center}
\vspace{.2in}
{\large {\bf Coloured Black Holes \\ \vspace{.2in}
in Higher Curvature String Gravity}} \\
\vspace{.4in}
{\bf P. Kanti }$^{(1)}$,
{\bf K. Tamvakis }$^{(2)\,,\,(3)}$ \\
\vspace{.3in}
{$^{(1)}$ \it Division of Theoretical Physics, Physics Department},\\
{\it University of Ioannina, Ioannina GR-451 10, GREECE}\\[4mm]
{$^{(2)}$ \it CERN, Theory Division},\\
{\it 1211 Geneva 23, SWITZERLAND}\\

\vspace{.5cm}
{\bf Abstract} \\
\vspace{.05in}
\end{center}
We consider the combined Yang Mills-Dilaton-Gravity
system in the presence of a Gauss-Bonnet term as it appears
in the $4D$ Effective Superstring Action.\,\,We give analytical
arguments and demonstrate numerically the existence of black 
hole solutions with non-trivial dilaton and Yang Mills hair
for the particular case of SU(2) gauge fields.\,\,The 
thermodynamical properties of the solutions are also discussed.\\[5mm]
PACS codes\,:\,04.20.Jb, 04.70.Bw, 11.15.Kc, 97.60.Lf
\vspace{0.2in}
%
\paragraph{}

\vspace{0.01in}
\begin{flushleft}
$^{(3)}$ on leave of absence from Physics Department, University
of Ioannina\\ 
\vspace{0.2in}
\,\,\,August 1996 \\
\end{flushleft}
\vspace{0.5in}
\hrule
\vspace{0.05in}
E-mails\,:\,pkanti@cc.uoi.gr, tamvakis@cc.uoi.gr
\end{titlepage}
\newpage

\begin{bf}1. Dilaton gravity with a Gauss-Bonnet term and an 
$SU(2)$ Yang Mills potential.\end{bf}

\paragraph{}
The effective theory of gravity resulting from String Theory
\cite{witten} at low
energies includes important modifications of Einstein's theory due to
the presence of the extra degrees of freedom such as dilatons, axions 
and Yang Mills fields. The loop-corrected Superstring Effective Action
through the contribution of the Gauss-Bonnet term leads to the 
existence of singularity-free cosmological solutions 
\cite{antoniadis}~\cite{easther} as well as to the existence of new
dilatonic black holes \cite{kanti}\,\cite{alexeyev}. The black hole 
solutions found possess hair \cite{bek} outside the horizon as
anticipated from general and perturbative in $\alpha'$ considerations
\cite{callan}\,\cite{mignemi}. The existence of black holes with
non-trivial hair has been demonstrated in the Einstein-Maxwell-Dilaton
system \cite{shapere}\,\cite{mitra} and the
Einstein-Yang Mills(-Dilaton-Axion)
system \cite{bizon}. In the present article we shall extend the analysis
of reference \cite{kanti} by including an SU(2) Yang Mills field.
We find new black hole solutions of the combined gravity-dilaton-SU(2) 
Yang Mills system in the (crucial) presence of the Gauss-Bonnet term.
\paragraph{}
Let us consider the heterotic string effective action, ignoring for
simplicity moduli and axion fields. Following the notation of
reference \cite{kanti}, we have
\begin{equation}
{\cal L}=-\frac{1}{2}R - \frac{1}{4} (\partial _\mu \phi )^2
+ \frac{\alpha '}{8g^2} e^{\phi }( R^2 _{GB}- F^{\alpha \mu \nu}
F_{\,\,\mu \nu} ^{\alpha})
\label{one}
\end{equation}
where $\alpha'$ is the Regge slope, $g^2$ is some gauge coupling 
constant and $F_{\mu\nu}$ is the Yang Mills field strength. The Gauss-
Bonnet term is defined as
\begin{equation}
R^2_{GB}=R_{\mu\nu\rho\sigma} R^{\mu\nu\rho\sigma}-
4 R_{\mu\nu} R^{\mu\nu} + R^2
\label{two}
\end{equation}
We shall consider only the contribution of an SU(2) Yang Mills field,
assuming trivial values for all other Yang Mills as well as ``matter"
fields.
\paragraph{}
At this point we shall make a spherically symmetric ansatz for the 
space-time metric
\begin{equation}
ds^2 = -e^\Gamma dt^2 + e^\Lambda dr^2 + r^2 (d\theta ^2 + sin^2 \theta
d\varphi^2)
\label{six}
\end{equation}
The functions $\Gamma$ and $\Lambda$ depend solely on the radius
$r$. For the SU(2) Yang Mills potential we shall adopt the magnetic
ansatz \cite{bizon}
\begin{equation}
{\bf{A}}=\frac{1+w}{e}\,[-\hat{\tau}_{\varphi} d\theta +
\hat{\tau}_{\theta} sin\theta d\varphi]
\label{poten}
\end{equation}
where $e$ is an SU(2) gauge coupling constant, $w=w(r)$ is a radial
function, and ($\hat{\tau}_r,\hat{\tau}_{\theta},
\hat{\tau}_{\varphi}$) is the anti-Hermitian SU(2) basis expressed in
polar coordinates~; e.g. $\hat{\tau}_r=\hat{\bf{r}} \cdot \tau$, and 
$[\hat{\tau}_a,\hat{\tau}_b]=\epsilon_{abc} \hat{\tau}_c$ with the
indices ranging over ($r,\theta,\varphi$). The field strength
corresponding to (\ref{poten}) is
\begin{equation}
{\bf{F}}=\frac{w'}{e}\,[-\hat{\tau}_{\varphi}\, dr \wedge d\theta +
\hat{\tau}_{\theta} \,sin\theta \,dr \wedge d\varphi]-
\frac{(1-w^2)}{e}\, \hat{\tau}_r\, sin\theta \,d\theta \wedge d\varphi
\label{fmn}
\end{equation}
%
\begin{bf}2. Some analytical considerations.\end{bf}
\paragraph{}
Using the static, spherically symmetric ansatz (\ref{six}) and
(\ref{fmn}), the dilaton and Yang-Mills equation as well as the $(tt)$,
$(rr)$ and $(\theta\theta)$ component of the Einstein's equations
resulting from (\ref{one}) take the form
\begin{eqnarray}
\phi''+\phi'(\frac{\Gamma'-\Lambda'}{2}+\frac{2}{r})&=&
\frac{\alpha'e^\phi}{g^2 r^2}\left(\Gamma'\Lambda'e^{-\Lambda}+
(1-e^{-\Lambda})[\Gamma''+
\frac{\Gamma'}{2}(\Gamma'-\Lambda')]\right. \nonumber\\ [3mm] 
&+& \left. \frac{w^{'2}}{e^2}+ \frac{(1-w^2)^2}{2e^2r^2}\,e^{\Lambda}\right)
\label{equ1} \\[3mm] 
w'' + w'\,(\frac{\Gamma'-\Lambda'}{2} + \phi') &+&
\frac{w(1-w^2)}{r^2}\,e^{\Lambda} =0
\label{equ2} \\[3mm]
\Lambda'\left(1+\frac {\alpha' e^\phi} {2 g^2 r} \phi'
(1-3 e^{-\Lambda})\right)
&=&\frac{r\phi'^2}{4}+\frac {1-e^\Lambda}{r}
+\frac{\alpha' e^\phi}{g^2 r}(\phi''+\phi'^2)(1-e^{-\Lambda})\nonumber 
\\ [3mm]
&+& \frac{\alpha' e^{\phi}}{2 g^2 r}\, \left( \frac{w^{'2}}{e^2}+ \frac{(1-w^2)^2}{2e^2r^2}\,e^{\Lambda}\right)
\label{equa3} \\[3mm]
\Gamma'\left(1+\frac {\alpha' e^\phi}{2 g^2 r}\phi'
(1-3 e^{-\Lambda})\right)
&=&\frac{r \phi'^2}{4}+\frac{e^\Lambda-1}{r}+ \frac{\alpha' e^{\phi}}
{2 g^2 r}\, \left( \frac{w^{'2}}{e^2}-\frac{(1-w^2)^2}{2e^2r^2}\,e^{\Lambda}\right)
\label{equ4} \\[3mm]
\Gamma''+\frac{\Gamma'}{2}(\Gamma'-\Lambda')+
\frac{\Gamma'-\Lambda'}{r}
&=&-\frac{\phi'^2}{2}+\frac{\alpha' e^{\phi-\Lambda}}{g^2 r}
\left(\phi'\Gamma''+(\phi''+\phi'^2)\Gamma'\right.
\nonumber \\[3mm]
&+&\left. \frac{\phi'\Gamma'}{2}(\Gamma'-3\Lambda')\right)
+ \frac{\alpha' e^{\phi}}{2 g^2 e^2 r^4}\, (1-w^2)^2\,e^{\Lambda} 
\label{equ5}
\end{eqnarray}
The corresponding components of $T_{\mu \nu}$ are
\begin{eqnarray}
T^t_t&=&- e^{-\Lambda} \frac {\phi'^2}{4}-
\frac {\alpha'e^{\phi-\Lambda}}{g^2 r^2} \left\{(\phi''+\phi'^2)
 (1-e^{-\Lambda}) - \frac{\phi' \Lambda'}{2} (1-3 e^{-\Lambda}) +
\frac{w'^2}{2e^2} + \frac{(1-w^2)^2 e^{\Lambda}}{4 e^2 r^2}\right\}
\nonumber \\ [3mm]
T_r^r&=& e^{-\Lambda} \frac {\phi'^2} {4}-
\frac {\alpha'} {g^2 r^2} e^{\phi-\Lambda}\left\{\frac{ \phi'
 \Gamma'}{2} (1-3 e^{-\Lambda}) -\frac{w'^2}{2e^2} + \frac{(1-w^2)^2 e^{\Lambda}}{4 e^2 r^2}\right\}
  \label{comp} \\ [3mm]
T_\theta^\theta&=&- e^{-\Lambda} \frac {\phi'^2}{4}
 +\frac {\alpha' e^{\phi-2 \Lambda}}{2 g^2 r}\left\{ 
 \Gamma'' \phi' + \Gamma' (\phi'' + \phi'^2) + \frac {\Gamma'
 \phi'} {2} (\Gamma'-3 \Lambda')+ \frac{(1-w^2)^2 e^{2 \Lambda}}
{2 e^2 r^3}\right\}\nonumber
\end{eqnarray}
Due to the presence of the higher curvature terms, the assumption
of positive definiteness of the time-component of the ``energy-momentum"
tensor breaks down. This point is crucial for the evasion of the
no-hair conjecture \cite{bek} and the existence of new black hole
solutions. 
\paragraph{}
Let us study now our system at infinity. Demanding asymptotically flat
solutions, we expand the functions $e^{\Lambda(r)}$, $e^{\Gamma(r)}$,
$\phi(r)$ and $w(r)$ in a power series in $1/r$ and substitute them
in the equations (\ref{equ1})~-~(\ref{equ5}). From eq.(\ref{equ2})
in the first order we take the constraint\,: $w_\infty\,(1-w^2_\infty)=0$.
As a result, the only two options allowed are either
$w_\infty=0$ or $w_\infty=\pm 1$. The first leads to $w_1=0 $ 
in the second order, to $w_2=0$ in the third order and
eventually to the vanishing of all $w_n$'s. This case corresponds
to an Abelian potential 
\begin{equation}
{\cal A} \sim \frac{1/e}{r} \Rightarrow B_r= F_{\theta\varphi} \sim
\frac{1/e}{r^2}
\end{equation}
with magnetic charge $1/e$. The other option, namely $w_\infty=\pm 1$,
leads to non-vanishing $w_n$'s
and corresponds to the non-Abelian sector of SU(2). In this case the 
radial magnetic field is $B_r \sim 1/r^3$ implying that the Yang Mills 
charge $Q_{YM}$ vanishes at infinity. We shall study the non-Abelian
case in this article and, thus, make the choice $w_\infty=\pm1$ which
leads us to the following asymptotic form of the solution near
infinity 
\begin{eqnarray}
e^{\Lambda(r)}&=& 1+ \frac{2 M}{r} + \frac{16 M^2 -D}{4 r^2} 
+ O(\frac{1}{r^3}) \label{inflam}\\[3mm]
e^{\Gamma(r)}&=& 1- \frac{2 M}{r} + O(\frac{1}{r^3}) 
\label{infgam}\\[3mm]
\phi(r)&=& \phi_{\infty} + \frac{D}{r} +  \frac{D M}{r^2} +
O(\frac{1}{r^3}) \label{infphi}\\ [3mm]
w(r)&= &\pm \left(1 +\frac{D_w}{r} + O(\frac{1}{r^2}) \right)
\label{infw}
\end{eqnarray}
$M$ is the ADM mass and $D$ is the dilaton charge defined over a
two-sphere at infinity as~\cite{mitra}
\begin{equation}
D=-\frac{1}{4\pi}\int d^2\Sigma^\mu \nabla_\mu \phi
\end{equation}
\paragraph{}
Let us now direct our attention to the other end of the allowed spatial
range, i.e. near the event horizon. We note that equation (\ref{equ4}) 
can be solved for $e^\Lambda$ as
\begin{equation}
e^\Lambda=\frac{-\beta+\delta\sqrt{\beta^2-4\alpha \gamma }}{2 \alpha}
\label{elambda}
\end{equation}
where $\delta=\pm1$ and
\begin{eqnarray}
\alpha&=&1- \frac{\alpha'e^{\phi}}{4 g^2 r^2}\,\frac{(1-w^2)^2}{e^2}
\quad, \quad \gamma=\frac{3 \alpha'e^\phi }{2 g^2}\Gamma'\phi'
\nonumber \\[2mm]
\beta&=&\frac{\phi^{'2} r^2}{4}-1-\Gamma'(r+\frac{\alpha' e^\phi\phi'}
{2 g^2}) + \frac{\alpha'e^{\phi}}{2 g^2}\, \frac{w^{'2}}{e^2}
\end{eqnarray}
As a result $e^{\Lambda}$, as well as $\Lambda'$, can be eliminated from equations (\ref{equ1}), (\ref{equ2}), (\ref{equa3}) and (\ref{equ5}).
Choosing any three of them, since only three of them are independent,
we obtain the system
\begin{eqnarray}
&\phi''&=-\frac{d_1}{d}
\label{thone} \\ [2mm]
&\Gamma''&=-\frac{d_2}{d}
\label{three}\\[2mm]
&w''&= -\frac{d_3}{d}
\label{thtwo} 
\end{eqnarray}
where $d$, $d_1$, $d_2$, $d_3$ are complicated functions of
$\phi'$, $\Gamma'$, $w'$, $\phi$, $w$, and $r$.
Demanding that $\phi''_h$ and
$w''_h$ are finite, we obtain from (\ref{thone}) and (\ref{thtwo}) the
constraints
\begin{eqnarray}
&~&\phi_h^{'2}\,[\,2 e^{2 \phi_h} C_h r_h \,(e^{2 \phi_h} + e^{\phi_h} r_h^2
+r_h^4)-8 e^{\phi_h} r_h^7\,] + \nonumber \\[3mm] &~& 
\phi'_h \,[ \,e^{4 \phi_h} C_h^2 + 4 e^{\phi_h} C_h r_h^2 \,(e^{2 \phi_h}
+ 2 e^{\phi_h} r_h^2 + r_h^4) -16 r_h^8\,] -  
\label{phicon} 
\\[3mm] &~& 
e^{3 \phi_h} C_h^2 r_h + 24 e^{2 \phi_h} C_h r_h^3 +
8 e^{\phi_h} C_h r_h^5 -48 e^{\phi_h} r_h^5 =0 \nonumber
\end{eqnarray}
and
\begin{equation}
w'_h = - \frac{ w_h\,(1-w_h^2)\,(1+e^{\phi_h} \phi_h' / 2r_h)}
{ r_h \,(1- e^{\phi_h} C_h /4 r_h^2)}
\label{wcons}
\end{equation}
where $C_h=(1-w_h^2)^2/e^2$. Since $\alpha'/g^2$ always multiplies
$e^\phi$, we may eliminate it from our calculations and restore it
in the end. Substituting these two into eq.(\ref{three}) gives
\begin{equation}
\Gamma''= -\Gamma'^2 + {\cal{O}}(1) \Rightarrow 
\Gamma'=\frac{1}{r-r_h} + {\cal{O}}(1)
\end{equation}
Expanding equation (\ref{elambda}) near the horizon in powers of
$\Gamma'$, for $\delta=+1$ we obtain
\begin{equation}
e^{\Lambda}= \frac{2 r^2 \,(2 r+ e^{\phi} \phi')}
{4 r^2 -e^{\phi}C}\, \Gamma' +{\cal{O}}(1) \label{delta1}
\end{equation}
while the choice $\delta=-1$ leads to $e^{\Lambda}={\cal{O}}(1)$ which 
is not a black hole solution. Taking into account all
the above, we may conclude that the unique black hole solution with 
$\phi'$, $\phi''$, $w'$, $w''$ finite and $\Gamma' \rightarrow \infty$
near the horizon has the expansion
\begin{eqnarray}
e^{\Gamma(r)} &=& \gamma_1 (r-r_h)+{\cal O}(r-r_h)^2 
\label{expgam} \\[3mm]
e^{-\Lambda(r)}&=& \lambda_1 (r-r_h)+{\cal O}(r-r_h)^2 
\label{explam}\\ [3mm]
\phi(r)&=& \phi_h + \phi'_h\,(r-r_h) + \frac{\phi''_h}{2}\,(r-r_h)^2 +
{\cal O}(r-r_h)^3  \label{expphi}\\[3mm]
w(r)&=& w_h + w'_h\,(r-r_h) + \frac{w''_h}{2}\,(r-r_h)^2 +{\cal O}(r-r_h)^3
\label{expw}
\end{eqnarray}
where $r_h$, $\phi_h$, $w_h$ and $\gamma_1$ are free parameters 
while from eq.(\ref{delta1}) we get 
\begin{equation}
\lambda_1=\frac{4 r_h^2 -e^{\phi_h}\,C_h}
{2 r_h^2 \,(2 r_h+ e^{\phi_h} \phi_h')}
\label{lamcon}
\end{equation}
\paragraph{}
Note that the constraint (\ref{phicon}) is an algebraic second order
equation  for $\phi'_h$ which has two real solutions, $\phi'_{\pm}$,
only if its discriminant is positive. This ultimately gives
\begin{equation}
(C_h-z_{+}) (C_h-z_{-}) \geq 0
\end{equation}
where 
\begin{equation}
z_{\pm}=\frac{4}{x^3}\,[-(2+3x+3x^2) \pm \sqrt{(2+3x+3x^2)^2+
6x^2-1}\,]
\label{zsol}
\end{equation}
having set $x=\alpha' e^{\phi_h} /g^2 r_h^2$. When
$x \leq 1/\sqrt{6}$, or
\begin{equation}
\frac{\alpha' e^{\phi_h}}{g^2} \leq \frac{r_h^2}{\sqrt{6}}
\label{crit1}
\end{equation}
the discriminant is always positive. This has the obvious interpretation 
that if the Gauss-Bonnet effective coupling $\alpha' e^{\phi_h}/g^2$ is 
smaller than the critical value $r_h^2/\sqrt{6}$, there is always a 
dilatonic black hole solution. The same constraint was derived in 
the case of Einstein-Dilaton-Gauss-Bonnet theory \cite{kanti}. 
There seems to be, 
however, an additional possibility in the present case. The inequality
(47) can be true either when $C_h \geq z_+>z_-$ or when
$ C_h \leq z_- < z_+$. Since, however, $z_-<0$ and $C_h>0$, the latter case
is impossible and only $ C_h \geq z_+$ could, perhaps, be realized. Then, for
$x \geq 1.11949 $, the inequality $0 < z_+<1$ is true. Later we shall
justify the choice $w_h < 1$, or equivalently $C_h<1$. Thus, 
solutions could also be present if the effective Gauss-Bonnet coupling is
in the region
$\frac{\alpha' e^{\phi_h}}{g^2} \geq 1.11949\,r_h^2$.
 Nevertheless, no solutions were found in this new region.
\paragraph{}
Before we proceed to describe our numerical procedure and finally plot
our solutions, let us compute analytically the temperature and
entropy of the black hole. Introducing the Euclidean version of our
metric
\begin{equation}
ds^2 = e^\Gamma d\tau^2 + e^\Lambda dr^2 + r^2 (d\theta ^2 + sin^2 \theta
d\varphi^2)
\end{equation}
where $\tau$ is a periodic coordinate ranging over the interval
$(0,\beta)$, we define the temperature as 
\cite{mignemi}~\cite{gibbons}
\begin{equation}
T= \frac{\kappa}{2 \pi}=\frac{1}{4\pi}\,\frac{1}{\sqrt{g_{tt} g_{rr}}}
\left(\frac{dg_{tt}}{dr} \right)_{r=r_h}=\,
\frac{\sqrt{\gamma_1 \lambda_1}}{4 \pi}
\label{temp1}
\end{equation}
where $\kappa$ is the surface gravity of the black hole. On the other hand
rearranging the equations of motion, we obtain
\begin{equation}
\frac{d}{dr}\left(e^{(\Gamma-\Lambda)/2} (\Gamma'-\phi') \,r^2 -
\frac{\alpha'e^\phi}{g^2}e^{(\Gamma-\Lambda)/2}
[(\phi'-\Gamma')\,(1-e^{-\Lambda})+ e^{-\Lambda} \phi' \Gamma' r]
\right)=0
\label{identi1}
\end{equation}
The equation (\ref{identi1}) is an identity.
Integrating over the interval $(r_h,r)$, we obtain
\begin{equation}
2M+D=\sqrt{\gamma_1 \lambda_1}\,(r_h^2 +\frac{\alpha'e^{\phi_h}}
{g^2})
\label{ident2}
\end{equation}
which gives
\begin{equation}
T=T_{SW}\,\frac{r_h\,(2M+D)}{(r_h^2 +\frac{\alpha'e^{\phi_h}}{g^2})}
\label{temp3}
\end{equation}
in terms of the temperature of a Schwarzschild black hole
$T_{SW}=1/8 \pi M = 1/4 \pi r_h$ and the asymptotic parameters
$M$ and $D$. Since $T$ is a function of positive constants, it never
vanishes. As a result, there is no mechanism to prevent the complete
evaporation of the black hole. However, this is indeed the case only when
the parameters $r_h$ and $\phi_h$ obey the constraint (\ref{crit1}).
 If, due to thermal evaporation, the parameters of
the black hole stop to obey the above constraint, then this
gravitational system can no longer be described by the concept of a
regular black hole.
\paragraph{}
The entropy $S$ can be derived from the free energy $F(\beta)$ of
the system as
\begin{equation}
S=\beta\,[ \, \frac{\partial (\beta F)}{\partial \beta}
-F\,]
\label{entro1}
\end{equation}
The free energy is defined as $F=I_E/\beta$, where $I_E$ is the 
Euclidean version of the action
\begin{equation}
I_E=\int d^4x \sqrt{g} \,{\cal{L}} - \int d^4x \sqrt{g}\, K
\label{action}
\end{equation}
where $K$ is a suitably subtracted boundary term that comes from
$R$ and $R^2_{GB}$.
We may perform the time and angular integration in (\ref{action}) and
by using the equations of motion the action takes the form
\begin{equation}
I_E=2 \pi \beta \,e^{(\Gamma-\Lambda)/2}\,[\,(\Gamma'-\phi')\,r^2 +
\frac{\alpha' e^\phi}{g^2}\,(1-e^{-\Lambda})\,(\Gamma'+\phi')-
\frac{\alpha' e^\phi}{g^2}\, r e^{-\Lambda} \Gamma' \phi'\,]
\,|^{\infty}_{r_h}
\end{equation}
Substituting again the expansions near the horizon as well as near
infinity and by making use of the definition (\ref{entro1}), we obtain 
\begin{equation}
S=\frac{A_H}{4}\,( 1+ \frac{\alpha' e^{\phi_h}}{g^2 r^2_h})
\label{entro2}
\end{equation}
where $A_H=4 \pi r^2_h$ is the area of the event horizon and 
$S_{SW}=A_H/4$ is the Bekenstein-Hawking formula \cite{beken} for
the entropy of the Schwarzschild black hole.
\paragraph{}
We note from eq.(\ref{one}) that in the limit $\phi \rightarrow
- \infty$ the effective coupling $\alpha'e^{\phi}/g^2$ vanishes. Then, 
the contribution of the Gauss-Bonnet term and the SU(2) potential to
the equations of motion becomes trivial. We can show \cite{kanti} that
in this case the  only acceptable solution is the standard Schwarzschild
one with constant dilaton field, in agreement with the no-hair theorem.
In this limit, we expect the temperature $T$ and the entropy $S$ of the
coloured black hole to approach the corresponding Schwarzschild ones,
$T_{SW}$ and $S_{SW}$. As we can see from (\ref{temp3}), since in this limit
$D \rightarrow 0$, and (\ref{entro2}), this is exactly what we obtain.\\[3mm]
\begin{bf}3. Numerical considerations and conclusions.\end{bf}
\paragraph{}
Taking into account the constraints (\ref{phicon}),(\ref{wcons}) and
(\ref{lamcon}) we may conclude that the parameters
of the problem are $r_h$, $\phi_h$, $w_h$  and  $\gamma_1$. Note that 
the equations of motion (\ref{equ1})-(\ref{equ5}) do not involve 
$\Gamma(r)$ but only $\Gamma'(r)$. The final integration determining
$\Gamma(r)$ involves the integration constant $\gamma_1$ which will
fixed by demanding asymptotic flatness through (\ref{infgam}).
\paragraph{}
Considering the Yang-Mills equation at the horizon, we obtain
\begin{equation}
(e^{-\lambda})'_h w'_h + \frac{w_h\,(1-w_h^2)}{r_h^2} = 0
\Rightarrow 
sign\,w'_h=sign\,w_h\,(w^2_h-1)
\end{equation}
According to this equation, if we choose the initial value
$w_h$ to be greater than one, then $w'_h>0$, which means that  
$w\,(r)$ increases with $r$. If we want to constrain $w\,(r)$ at
infinity by $w\,(\infty)=\pm1$ then, a local maximum must occur
at some point $r=r_s$. At this point we would have $w\,(r_s) > 1$ and 
$w'(r_s) =0$. Using again the Yang-Mills equation, we obtain
\begin{equation}
sign\,w''(r_s)= sign\,w\,(r_s)\,[w^2(r_s)-1]
\end{equation}
If $w\,(r_s) > 1$, then $w''(r_s) > 0$, which means
that we can have only a local minimum at $r_s$. As a result, the initial 
value of $w\,(r)$ must fall inside the interval $(-1,1)$. Moreover,
we observe that the equations of motion remain invariant under the
transformation $ w \rightarrow -w$. Thus, it would be sufficient to
choose initial values of $w_h$ in the interval $(0,1)$.
\paragraph{}
The equations of motion (\ref{equ1})-(\ref{equ5}) are invariant under
the combined transformation
\begin{equation}
\phi \rightarrow \phi+\phi_0 \quad , \quad
r \rightarrow r \,e^{\phi_0/2} \nonumber
\end{equation}
As a result, it is sufficient to vary only one of $r_h$ and $\phi_h$.
Furthermore, we may use the above invariance to set a unique
mass scale for all solutions by imposing the asymptotic condition
$\phi_{\infty}=0$. This requires a shift $\phi \rightarrow \phi-
\phi_{\infty}$ accompanied by a rescaling $r \rightarrow r 
e^{-\phi_{\infty}/2}$. Since the radial coordinate has been rescaled,
the other two asymptotic parameters, $M$ and $D$, are also rescaled
according to the rule $M \rightarrow M e^{-\phi_{\infty}/2}$ and
$D \rightarrow D e^{-\phi_{\infty}/2}$. Similarly, the temperature $T$
is also rescaled as $T \rightarrow T e^{\phi_{\infty}/2}$.
\paragraph{}
In order to perform the numerical integration it is convenient to 
set $\frac{\alpha'}{g^2}=e=1$. We fix the value of the horizon at 
$r_h=1$ and start by giving initial values to the remaining parameters 
$\phi_h$ and $w_h$. Starting from the expansions (\ref{expgam})-(\ref{expw}),
at $r=r_h+\epsilon$ with $\epsilon\simeq O(10^{-8})$, we integrate the 
system (\ref{thone})-(\ref{three}) towards $r \rightarrow \infty$ using 
the fourth order Runge-Kutta method with an automatic step procedure
and accuracy $10^{-8}$. The integration stops when the flat
space-time asymptotic limit (\ref{inflam})-(\ref{infw}) is reached.
In the second allowed region of the Gauss-Bonnet effective coupling,
defined by (47), and for the choices 
$\phi'_h=\phi'_{+}$ and $\phi'_h=\phi'_{-}$ we found solutions with regular
and non-regular behaviour respectively near the horizon. However, in
both cases the solutions lacked the right asymptotic behaviour 
(\ref{inflam})-(\ref{infw}) near infinity.
\paragraph{}
Making the choice $\phi'_h=\phi'_{-}$, which corresponds to the choice
$\phi'_h=\phi'_{+}$ for the Einstein-Dilaton-Gauss-Bonnet theory, we
were able to find regular asymptotically flat black hole solutions in
the first allowed region defined by (\ref{crit1}).
For every initial value of the shooting parameter $\phi_h$ there is a 
discrete family of initial values of $w_h$ which results into a discrete
family of asymptotically flat black hole solutions characterized by the
number $n$ of zeros (nodes) of $w(r)$. We find that as $\phi_h ~\rightarrow~
-\infty$ the variation of the dilaton field $\phi(r)$ with $r$ becomes
weaker and the dilaton eventually behaves as a constant. The dilaton charge
$D$ moves towards zero and the horizon takes on its Schwarzschild value
$r_h=2 M$. As we expect, the temperature $T$ and the entropy $S$ also
take on the corresponding Schwarzschild ones. In the Table we display
three sets of corresponding values of $\phi_h$, $w_h$, and $\phi_\infty$.
The displayed values of $r_h$, $2M$, $D$, and $T$  are the rescaled ones
corresponding to imposing the condition $\phi_{\infty}=0$ at the
end of our computation.
Plots involving the dilaton field $\phi(r)$, for three different coloured
black hole solutions, after the imposition of the asymptotic condition,
are given in Figure 1. Figure 2 
depicts the Yang-Mills function $w$ of the $n=1,2,3$ coloured black hole
solutions as well as the metric function $g_{tt}(r)$. The dependence of
the other metric function $g_{rr}(r)$ is presented in Figure 3.
\newpage
\begin{center}
{\bf TABLE I}\\[3mm]
Parameters of solutions for $r_h=1$\\[4mm]
$\begin{array}{|c|c|c|c|c|c|c|c|}\hline
\hspace{0.3cm} \phi_h \hspace{0.3cm} & \hspace{0.2cm} n \hspace{0.2cm}
& \hspace{0.9cm} w_h \hspace{0.9cm} & \hspace{0.4cm} \phi_{\infty}
\hspace{0.4cm} & \hspace{0.4cm} r_h \hspace{0.4cm} &
\hspace{0.4cm} 2M \hspace{0.4cm} &  \hspace{0.4cm} D \hspace{0.4cm}
& \hspace{0.4cm} T \hspace{0.4cm}\\ \hline
-1.0 & 1 & 0.255948298 &-1.43634 & 2.05067 & 2.28110 & 0.30503
& 0.03578 \\
      & 2 & 0.042444284 &         &         &   &   &    \\
      & 3 & 0.006916638 &         &         &   &   &    \\ \hline
-2.5 & 1 & 0.262218961 &-2.60493 & 3.67836 & 3.76249 & 0.14371 
& 0.02123 \\
      & 2 & 0.043590755 &         &         &   &   &    \\
      & 3 & 0.007101253 &         &         &   &   &    \\ \hline
-5.0 & 1 & 0.265648181 &-5.00906 & 12.2378 & 12.2629 & 0.04275 
& 0.00650 \\
      & 2 & 0.044212029 &         &         &   &   &    \\
      & 3 & 0.007205569 &         &         &   &   &    \\ \hline
\end{array}$ \\ [10mm]
\end{center}
\paragraph{}
In the same region (\ref{crit1}) and for the choice $\phi'_h=\phi'_{+}$
we found solutions which had the right
asymptotic behaviour (\ref{inflam})-(\ref{infw}) near infinity  and seemed
to comprise a second group of regular black hole solutions. Examining more
closely their behaviour near the horizon, they were found to possess a much
more complicated structure. This structure is characterized mainly by the
behaviour of the metric component $g_{rr}$, which is also given in Figure 3,
while the other component, $g_{tt}$, exhibits a typical black hole behaviour.
As we move from infinity towards the origin, the solution firstly reaches
the value $r=r_s$, around which we may write the expansions
\begin{eqnarray}
e^{-\Lambda}&=&\lambda_s + \lambda_2\,(r-r_s)^2 + ... \\[3mm]
\Gamma'&=&\gamma_1 + \gamma_2\,(r-r_s) + ... \\[3mm]
\phi&=& \phi_s + \phi'_s (r-r_s) + \frac{\phi''_s}{2} (r-r_s)^2 + ...\\[3mm]
w&=& w_s + w'_s (r-r_s) + \frac{w''_s}{2} (r-r_s)^2 + ...
\label{sols}
\end{eqnarray}
If we substitute these expansions into the equations of motion we obtain
a set of constraints which determine $\gamma_2$, $\phi'_s$, $w'_s$, $\phi''_s$, and $w''_s$ as functions of the free parameters $\lambda_s$, $\lambda_2$, $\gamma_1$, $\phi_s$, and $w_s$. The {\it {curvature invariant}} $R_{\mu \nu \rho\sigma} R^{\mu \nu \rho\sigma}$ near $r_s$ was found to be 
\begin{equation}
R_{\mu \nu \rho\sigma} R^{\mu \nu \rho\sigma}= \lambda_s^2 \,(\gamma_2 +
\frac{\gamma_1^2}{2})^2 + \frac{2\gamma_1^2 \lambda_s^2}{r_s^2} +
\frac{4\,(1-\lambda_s)^2}{r_s^4} + {\cal{O}}(r-r_s)
\end{equation}
We come to the conclusion that the surface $r=r_s$ is a regular one whose
existence may be interpreted as an unsuccessful attempt of nature to 
form a regular horizon. The significance of this fact will become clear
below. Moving beyond the value $r=r_s$, the solution
eventually stops at the value $r=r_x$. The asymptotic behaviour of the
fields near $r_x$ is
\begin{eqnarray}
e^{-\Lambda}&=& \lambda_x + \lambda_1 \sqrt{r-r_x} + ...\\[3mm]
\Gamma'&=& \gamma_1+ \gamma_2 \sqrt{r-r_x} + ...\\[3mm]
\phi&=& \phi_x + \phi'_x (r-r_x) + \phi''_x\,(r-r_s)^{3/2} + ...\\[3mm]
w&=& w_x + w'_x (r-r_x) + w''_x (r-r_s)^{3/2} + ...
\label{solx}
\end{eqnarray}
where again $\gamma_2$, $\phi'_x$, $w'_x$, $\phi''_x$ and
$w''_x$ are given through the equations of motion as functions of the free parameters $\lambda_x$, $\lambda_1$, $\gamma_1$, $\phi_x$ and $w_x$. The {\it {curvature invariant}} $R_{\mu \nu \rho\sigma} R^{\mu \nu \rho\sigma}$
near $r_x$ takes the form
\begin{equation}
R_{\mu \nu \rho \sigma} R^{\mu \nu \rho \sigma}= \left( \frac{\lambda_x^2}
{4}\,(\gamma_2 + \frac{  \gamma_1 \lambda_1}{2 \lambda_x})^2 +
\frac{\lambda_1^2}{2 r_x^2}\right) \frac{1}{r-r_x} + {\cal{O}}
(\frac{1}{\sqrt{r-r_x}}) \rightarrow \infty
\end{equation}
This means that at the value $r=r_x$ the solution ends up in a pure scalar
singularity. Since no regular horizon exists at some value $r=r_h>r_x$,
we may conclude that a naked singularity has been formed at $r_x$.
\paragraph{}
For the critical value of $\phi_h$, defined by (\ref{crit1}), the two
solutions $\phi'_{-}$ and $\phi'_{+}$ coincide, which means that the branch
of regular black hole solutions meets the branch of the solutions that
describe a naked singularity. 
As long as the parameters $r_h$ and $\phi_h$ of the solution obey the
constraint (\ref{crit1}), the only acceptable solution of the equations
of motion is the regular black hole solution (\ref{expgam})-(\ref{expw})
with the flat asymptotic behaviour (\ref{inflam})-(\ref{infw}) near infinity.
When $r_h$ and $\phi_h$ become such - maybe due to thermal evaporation -
that the constraint (\ref{crit1}) is violated ,
the expansions (\ref{expgam})-(\ref{expw}) break down and the solution
can no longer be described by the concept of the regular black hole. Then,
the system shifts to the other branch of solutions, described above,
corresponding to a naked singularity with exactly the same asymptotic
behaviour at infinity.
\paragraph{}
Alexeyev et al.\cite{alexeyev}, by using a method based on integrating over
an additional parameter, were able to examine the structure of the black hole solutions found in the Einstein-Dilaton-Gauss-Bonnet theory \cite{kanti}
inside the event horizon. According to their results, the solution under the
regular horizon $r_h$ exist only until the value $r=r_s$, where a pure scalar
singularity exists. Another solution begins from $r_s$ which exists only
until the ``singular horizon" $r_x$. When the Gauss Bonnet effective
coupling becomes larger, the distance between $r_s$ and $r_h$ becomes
smaller. Once the critical limit (\ref{crit1}) is reached, $r_h=r_s=r_x$
and the internal structure vanishes. Since the Yang-Mills function $w$
always resembles the behaviour of the dilaton field, we expect that the inclusion of the SU(2) Yang-Mills potential respects the above internal
structure of the regular black hole. Actually, the similarity between the
internal structure of the black hole found by Alexeyev et al. and the
structure of the solution describing a naked singularity found above is
obvious. However, the interpretation of $r_x$ and $r_s$ is different:
while in the first case they play the role of a ``singular horizon"
and a pure scalar singularity respectively inside the horizon, in the
second case they stand for a scalar singularity and an unsuccessful event
horizon respectively. As a result, we can make the further conjecture
that as the effective coupling becomes larger the internal structure
moves towards the horizon, for the critical value the points $r_x$,
$r_s$ and $r_h$ merge, while for values beyond the critical point
this structure penetrates the horizon and manifests itself as a naked
singularity.
\paragraph{}
{\bf Note added in proof.} While the present article was being completed,
a related article titled ``Dilatonic Black Holes with Gauss-Bonnet Term"
by T. Torii, H.~Yajima and K. Maeda, gr-qc/9606034 came into our
attention.
\paragraph{}
{\bf Acknowledgements}. We wish to thank J. Rizos and N. Mavromatos
for many helpful discussions.
\newpage
\begin{figure}
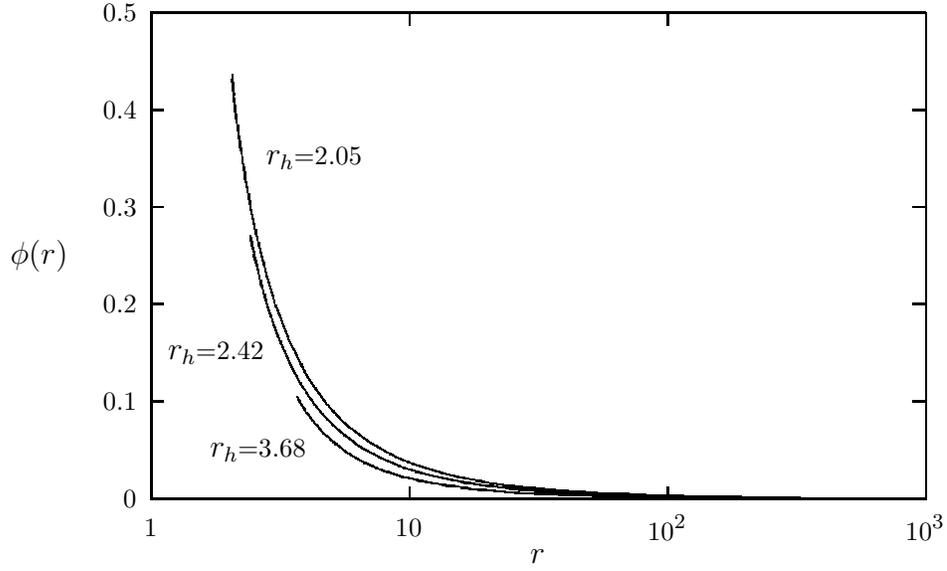

\begin{center}
\setlength{\unitlength}{0.240900pt}
\ifx\plotpoint\undefined\newsavebox{\plotpoint}\fi
\sbox{\plotpoint}{\rule[-0.200pt]{0.400pt}{0.400pt}}%

\caption{Dilaton field for coloured black hole solutions. Each curve
corresponds to a different solution characterized by a different
initial value of $\phi_h$.}
\end{center}
\end{figure}
\begin{figure}
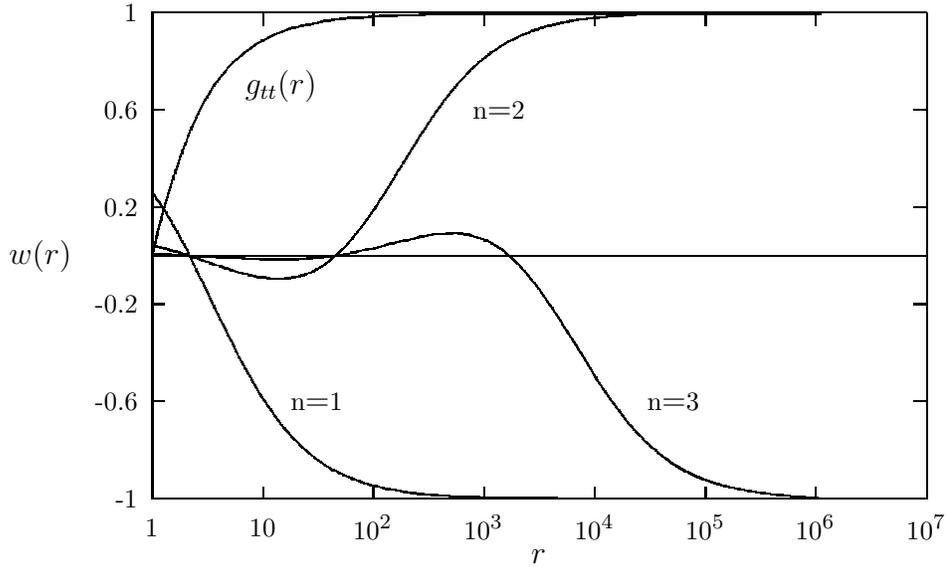

\begin{center}
\setlength{\unitlength}{0.240900pt}
\ifx\plotpoint\undefined\newsavebox{\plotpoint}\fi
\sbox{\plotpoint}{\rule[-0.200pt]{0.400pt}{0.400pt}}%
%
\caption{The Yang Mills function w of the $n=1,2,3$ coloured black
hole solutions as well as the metric component $g_{tt}$ for $r_h=1$ and $\phi_h=-1$.}
\end{center}
\end{figure}
\newpage
\begin{figure}
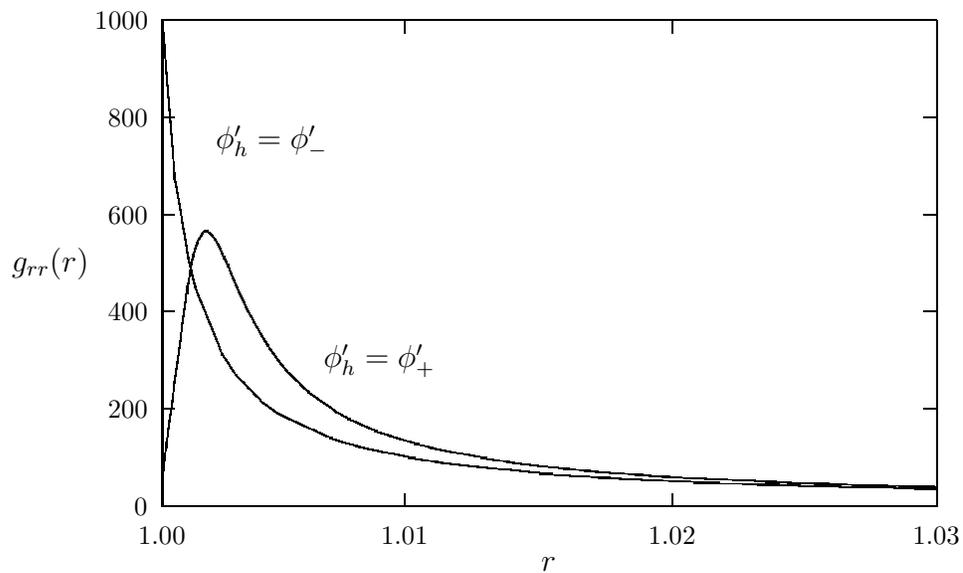

\begin{center}
\setlength{\unitlength}{0.240900pt}
\ifx\plotpoint\undefined\newsavebox{\plotpoint}\fi
\sbox{\plotpoint}{\rule[-0.200pt]{0.400pt}{0.400pt}}%
%
\caption{Dependence of the metric component $g_{rr}$ for $r_h=1$ and
$\phi_h=-2$. While for $\phi'_h=\phi'_{-}$ \, $g_{rr}$ moves towards an
infinite value leading to the formation of an event horizon,
for $\phi'_h=\phi'_{+}$ it reaches a maximum value at $r_s$ and then
declines leaving unshielded a scalar singularity.}
\end{center}
\end{figure}
\end{document}